\documentclass[tightenlines,floats,aps,nofootinbib,prd,showpacs]{revtex4}

\usepackage{graphicx,amsmath,amssymb}

\newcommand{\kk}{\mbox{\boldmath$k$}}
\newcommand{\xx}{\mbox{\boldmath$x$}}
\newcommand{\qq}{\mbox{\boldmath$q$}}
\newcommand{\vv}{\mbox{\boldmath$v$}}

\newcommand{\uk}{h\,{\rm Mpc}^{-1}}

\begin{document}
%
\title{Chasing the non-linear evolution of matter power 
  spectrum with numerical resummation method:  
  solution of closure equations} 
%
\author{Takashi Hiramatsu$^1$ and Atsushi Taruya$^{2,3}$}
\affiliation{$^1$Institute for Cosmic Ray Research, The University of
Tokyo, Kashiwa, Chiba 277-8582, Japan}
\affiliation{$^2$Research Center for the Early Universe, 
  School of Science, The University of Tokyo, Bunkyo-ku, Tokyo 113-0033, Japan}
\affiliation{$^3$Institute for the Physics and Mathematics of the Universe, 
  The University of Tokyo, Kashiwa, Chiba 277-8568, Japan}

\begin{abstract}
We present a new numerical scheme to treat the non-linear evolution of
cosmological power spectra. Governing equations for matter power spectra
have been previously derived by a non-perturbative technique with closure
approximation. Solutions of the resultant closure equations just
correspond to the resummation of an infinite class of perturbation
corrections, and they consistently reproduce the one-loop results of
standard perturbation theory. We develop a numerical algorithm to
solve closure evolutions in both perturbative and non-perturbative regimes.
The present numerical scheme is particularly suited for
examining non-linear matter power spectrum in general cosmological
models, including modified theory of gravity. As a demonstration, we
study weakly non-linear evolution of power spectrum in a class of
modified gravity models, as well as various dark energy models. 
\end{abstract}
\maketitle

\section{Introduction}
\label{sec:intro}

In the last decade, the late-time cosmic acceleration 
has been one of the most important discoveries 
in physics and cosmology (e.g., Refs.~\cite{Perlmutter1998, WMAP1}). 
Although the origin of late-time acceleration is thought to be 
a mysterious energy component called dark energy, 
a possibility of long-distance modification of general relativity is 
still viable (e.g., Refs.~\cite{DGP, Sotiriou2008}), and our understanding of
the nature of late-time acceleration is still lacking. Currently, the
dark energy 
equation-of-state parameter $w_{\rm de}$, which is phenomenologically 
introduced to characterize the cosmic acceleration and is 
defined as the ratio of the pressure to the energy density of dark energy,  
is consistent with a cosmological constant ($w_{\rm de}=-1$) at a level 
of $10\%$ precision, and with no evidence for time dependence
(e.g., Refs.~\cite{Tegmark2006, Komatsu2008}).  
Toward a deeper understanding of the nature of 
late-time acceleration, a precise measurement of both the cosmic expansion 
history and the growth of cosmic structure is a key to 
to distinguish between different models of dark energy, 
as well as to discriminate the dark energy from 
the modification of gravity.

Among several observational techniques, baryon acoustic oscillations 
imprinted on matter power spectrum and cosmic shear, measured from 
galaxy samples, are the most promising techniques sensitive to 
the expansion history and growth of structure. A crucial remark is 
that they strongly rely on the accurate predictions of non-linear 
matter power spectrum. Hence, in addition to the precise measurement, 
a high-precision theoretical template for the non-linear power 
spectrum must be developed in order to achieve order-of-magnitude 
improvement of the current constraints.

Recently, several analytical approaches to predict the non-linear 
power spectrum have been developed, complementary to the N-body 
simulations \cite{Crocce2005A, Crocce2005B,
Valageas2007A, Matarrese2007, Matsubara2007, Crocce2007, Taruya2008}. 
In contrast to the standard analytical calculation with 
perturbation theory (for a review, see Ref.~\cite{Bernardeau2002}), these
have been formulated in a non-perturbative  
way with techniques resumming a class of infinite series of higher-order 
corrections in perturbative calculation. 
Thanks to its non-perturbative formulation, the applicable 
range of the prediction has been greatly improved,  
and the non-linear evolution of baryon acoustic oscillations was 
found to be accurately described with a percent-level precision.

Note, however, that these analytical calculations involve several 
approximations or simplifications in order to make the analysis 
tractable. This severely limits the applicable range and/or  
the versatility of predictions. For example, in 
Refs.~\cite{Crocce2007,Taruya2008}, 
a perturbative treatment called Born approximation has been 
partly adopted in order to evaluate the non-perturbative 
expressions for power spectrum. 
Furthermore, most of the analysis presented so far rely on 
the Einstein-de Sitter approximation, 
in which all the calculations done in the Einstein-de Sitter universe 
are extended to apply to the other cosmological model by simply 
replacing the linear growth factor in Einstein-de Sitter universe with 
that in the other cosmology (see Sec.~\ref{subsec:1loopPT_in_DE} in 
detail). This is very crucial in studying the 
non-linear matter power spectrum in general cosmological models, 
especially in modified gravity models.

In the present paper, in order to bring out the advantage of 
non-perturbative formulation as much as possible, we present a numerical 
resummation scheme to calculate the non-linear matter power spectrum. 
Our treatment relies on the formalism developed by 
Ref.~\cite{Taruya2008}, in which the non-linear statistical method 
used in the subject of turbulence (e.g., Ref.~\cite{Kida1997})
was applied to the derivation of 
governing equations for power spectrum. The resultant equations called 
closure equations are the non-linear integro-differential equations 
coupled with non-linear propagator. The solution of closure equations 
effectively contains the information of the higher-order corrections, 
similar to the renormalized perturbation theory by Crocce \& 
Scoccimarro \cite{Crocce2005A, Crocce2005B,Crocce2007}.  
It has been shown that the analytical predictions based on the 
leading-order Born approximation agree with N-body simulations very well 
in a mildly non-linear regime, and a percent-level precision was 
achieved at some ranges \cite{Nishimichi2008}. 
The agreement of the prediction is further 
improved if taking account of the next-to-leading order correction 
\cite{Taruya2009}. Hence, with the numerical implementation of the 
closure equations, all orders of Born approximation are included, 
and the prediction will be much better than the analytical treatment. 
Further, the numerical treatment is particularly suited for 
studying the non-linear power spectrum in various cosmologies  
where the analytical calculations with Einstein-de Sitter approximation 
is no longer possible.

The paper is organized as follows. In Secs.~\ref{sec:preliminaries}
and \ref{sec:closure}, we 
briefly review the basic treatment of our approach and formalism.  
We then discuss how to solve closure equations in Sec.~\ref{sec:numerical}. 
As shown in Ref.~\cite{Taruya2008}, the closure equations automatically 
reproduce the leading-order results of 
standard perturbation theory if replacing the quantities in non-linear 
terms with linear-order ones. This treatment has been used for computing 
quasi non-linear spectrum in modified gravity models 
in Ref.~\cite{Koyama2009}. 
In Sec.~\ref{sec:results}, 
we present numerical solutions of closure equations in both full
non-linear and perturbative treatment and demonstrate how the present
scheme can treat analytically intractable cases.
Finally,  Sec.~\ref{sec:conclusion} is devoted to the discussion 
and conclusion.

\section{Preliminaries}
\label{sec:preliminaries}

Throughout the paper, 
we consider the evolution of mass distribution in the flat universe, 
neglecting the tiny contributions from the massive neutrinos. 
We treat the cold dark matter (CDM) plus baryon system as a pressureless
perfect fluid. 
Then, assuming the irrotationality of fluid flow, the governing equations 
for matter distribution become the continuity equation and the 
Euler equation coupled to the Newton potential $\phi$ (e.g.,
Ref.~\cite{Bernardeau2002}) : 
%
\begin{eqnarray}
 &&\frac{\partial\delta(\tau,\xx)}{\partial \tau} 
    + \theta(\tau,\xx)
    = -\frac{1}{aH}\nabla\cdot(\delta\,\vv), 
\label{eq:cont1} 
\\
 &&\frac{\partial \theta(\tau,\xx)}{\partial \tau} 
  + \left(2+\frac{d\ln H}{d\tau}\right)\theta(\tau,\xx) 
  = -\frac{1}{a^2H^2}\triangle\phi(\tau,\xx) 
  - \frac{1}{a^2H^2}\nabla\cdot(\vv\cdot\nabla\vv),
 \label{eq:euler1}
\end{eqnarray}
%
where $\delta$ is the mass density field, and $\theta$ is the 
velocity divergence defined as $\theta \equiv \nabla\cdot\vv/(aH)$. 
Here, we introduce the time variable given by $\tau=\log (a/a_0)$, with 
$a_0$ being the scale factor at the present time. With this time variable, 
the flat Friedmann equation becomes 
%
\begin{equation}
 H^2 = H_0^2 \left\{\Omega_{\rm m}\,e^{-3\tau} + 
\Omega_{\rm de}\,\exp\Bigl[-3\int^{\tau}_0 d\tau' \{1+w_{\rm de}(\tau')\} \Bigr]
\right\}.
\label{eq:sFriedmann}
\end{equation}
%
The quantity $H_0$ is the Hubble parameter at the present time, and 
$\Omega_{\rm m}$ and $\Omega_{\rm de}$ are 
the density parameters of the matter and dark energy, respectively.

To treat the non-linear evolution of matter power spectrum, we 
will work with the Fourier transform of the fluid equations,  
(\ref{eq:cont1}) and (\ref{eq:euler1}). They are given by 
%
\begin{eqnarray}
&&\frac{\partial \delta(\kk,\tau)}{\partial\tau} + \theta(\kk,\tau)
=-\int\frac{d^3\kk_1d^3\kk_2}{(2\pi)^3}\delta_{\rm D}(\kk-\kk_1-\kk_2)
\left\{1+\frac{\kk_1\cdot\kk_2}{|\kk_1|^2}\right\}\theta(\kk_1,\tau)
\delta(\kk_2,\tau),
\label{eq:Fourier_cont}
\\
&&\frac{\partial \theta(\kk,\tau)}{\partial\tau} + 
\left(2+\frac{d\ln H}{d\tau}\right)\theta(\kk,\tau)-
\left(\frac{k}{aH}\right)^2 \phi(\kk,\tau)
\nonumber
\\
&&\quad\quad\quad\quad\quad\quad\quad
=-\int\frac{d^3\kk_1d^3\kk_2}{(2\pi)^3}
\delta_{\rm D}(\kk-\kk_1-\kk_2)
\frac{(\kk_1\cdot\kk_2)|\kk_1+\kk_2|^2}{2|\kk_1|^2|\kk_2|^2}\theta(\kk_1,\tau)
\theta(\kk_2,\tau).
\label{eq:Fourier_euler}
\end{eqnarray}
%
As for the Poisson equation, we have 
%
\begin{equation}
-\frac{k^2}{a^2}\phi(\kk,\tau) = 
4\pi \,G_{\rm eff}(\kk,\tau)\rho_{\rm m}\,\delta(\kk,\tau).
  \label{eq:spoisson}
\end{equation}
%
Here, $G_{\rm eff}$ is the effective Newton constant, which generically 
depends on the scale and time in modified theory of 
gravity. In principle, the Newton potential can be a non-linear 
function of the density field. In fact, successful modified gravity models 
that explain late-time acceleration such as 
the Dvali-Gabadadze-Porrati (DGP) model \cite{DGP} and $f(R)$ gravity
models (for a review, see Ref.~\cite{Sotiriou2008}) have non-linear
interaction terms, which are essential to recover 
the general relativity on small scales \cite{Khoury2004}. 
In the present paper, we restrict our analysis to the cases with 
linear Poisson equation. The extension to the non-linear case is 
straightforward and is discussed in a separate paper \cite{Koyama2009}.

The evolution equations (\ref{eq:Fourier_cont}), 
(\ref{eq:Fourier_euler}) and (\ref{eq:spoisson}) 
can be further reduced to a compact form if we introduce the following quantity:
%
\begin{equation}
 \Phi_a(\kk,\tau) =\Bigl(
 \delta(\kk,\tau), \quad -\theta(\kk, \tau)\Bigr)\,; 
\quad (a=1,2). 
\label{eq:def_Phi}
\end{equation}
%
Then, we write down the evolution equations in a single form as 
%
\begin{equation}
\widehat{\Lambda}_{ab}\Phi_b(\tau,\kk) = \iint
    \frac{d^3\kk_1d^3\kk_2}{(2\pi)^6}\delta_D(\kk_1+\kk_2-\kk) 
    \gamma_{acd}(\kk_1,\kk_2)
     \Phi_c(\kk_1)\Phi_d(\kk_2), 
\label{eq:dPhi}
\end{equation}
%
where $\gamma_{acd}$ is the vertex function defined as
%
\begin{equation}
\begin{aligned}
\gamma_{112}(\kk_2,\kk_1) &= \gamma_{121}(\kk_1,\kk_2) 
        = \frac{1}{2}\left(1+\frac{\kk_1\cdot\kk_2}{|\kk_1|^2}\right),  \\
\gamma_{222}(\kk_1,\kk_2) &= \frac{1}{2}\left(\frac{|\kk_1+\kk_2|^2\kk_1\cdot\kk_2}{|\kk_1|^2|\kk_2|^2}\right).
\end{aligned}\label{eq:def_gamma}
\end{equation}
%
The operator $\widehat{\Lambda}_{ab}$ is defined by
%
\begin{equation}
\widehat{\Lambda}_{ab}=\delta_{ab}\frac{\partial}{\partial\tau} 
   + \Omega_{ab}(k,\,\tau),
\end{equation}
%
with the matrix $\Omega_{ab}$ being 
%
\begin{equation}
 \Omega_{ab}(k,\,\tau) =
\begin{pmatrix}
  0 &\,\,& -1 \\
{\displaystyle -4\pi G_{\rm eff}\,\frac{\rho_{\rm m}}{H^2} }&\,\,& 
{\displaystyle 2+\frac{d\ln H}{d\tau}}
\end{pmatrix}. \label{eq:def_Omega}
\end{equation}
%

\section{Closure equations}
\label{sec:closure}

In the present paper, we are especially concerned with the 
non-linear evolution of power spectrum defined by 
%
\begin{equation}
\langle\Phi_a(\kk,\tau)\Phi_b(\kk',\tau)\rangle 
      = (2\pi)^3\delta_D(\kk+\kk')P_{ab}(|\kk|;\tau), \label{eq:def_P}
\end{equation}
%
where the bracket $\langle\cdot\rangle$ stands for ensemble average. 
In the above definition, we have the three different power spectra,   
$P_{11}$, $P_{12}=P_{21}$, and $P_{22}$, which respectively correspond to 
$P_{\delta\delta}$, $-P_{\delta\theta}$ and $P_{\theta\theta}$.

For the analytical calculation of the power spectrum, 
the standard treatment of perturbation theory 
is to expand the quantity $\Phi_a$ as 
$\Phi_a=\Phi_a^{(1)}+\Phi_a^{(2)}+\cdots$, and to 
iteratively obtain the solutions $\Phi^{(n)}$ from Eq.~(\ref{eq:dPhi}). 
Substituting the perturbative solutions into the definition 
(\ref{eq:def_P}), we obtain the non-linear corrections to 
the power spectrum. This treatment is straightforward, but is
plagued by a poor convergence of the perturbative
expansion. Because of this, the applicable range of 
the standard perturbation theory (SPT) is restricted to a narrow range 
on large scales.

Recently, the improved treatment of the perturbation theory 
has been proposed by several authors employing the so-called
renormalized/resummation techniques \cite{Valageas2004A, Crocce2005A,
Crocce2005B, Valageas2007A, Matsubara2007, Matarrese2007, Izumi2007,
Crocce2007,Taruya2008,Pietroni2008}. 
In these treatments, the naive expansion of the SPT is 
re-organized by introducing 
the non-perturbative statistical quantities, and the information of the 
higher-order corrections in SPT is effectively incorporated into each 
order of expansions. As a result, even truncating the expansion 
at some orders still contains the non-perturbative effects of non-linear 
clustering, leading to the improvement of the convergence properties.

Here, among several non-perturbative techniques, 
we consider the closure theory proposed by 
Ref.~\cite{Taruya2008}, in which we have applied 
the non-linear statistical method 
commonly used in the subject of turbulence (e.g., Ref.~\cite{Kida1997}) to 
the cosmological perturbation theory. 
In this treatment, the renormalized expansion 
has been first constructed according to the renormalized perturbation 
theory by Ref.~\cite{Crocce2005A}. Then, we truncate the  
expansions at the one-loop order. Under the tree-level approximation 
of the vertex function, this leads to a closed system of the power 
spectrum and non-linear propagator. Though some non-perturbative 
properties are missed in this treatment, an advantage of this 
formulation is that we can compute the power spectrum 
numerically by solving the evolution equations, keeping full 
non-perturbative information of the non-linear clustering at 
the one-loop order. This forward modeling may be suitable for a 
fast computation of the power spectrum, 
unlike the backward treatment of the perturbative expansions, 
which requires the time-consuming multi-dimensional integrations.

Let us define the non-linear propagator, $G_{ab}(\kk|\tau,\tau')$, and 
the cross power spectra between different times, $R_{ab}(\kk;\tau,\tau')$: 
%
\begin{align}
 \left\langle\frac{\delta\Phi_a(\kk,\tau)}{\delta\Phi_b(\kk',\tau')}\right\rangle
 &= G_{ab}(\kk|\tau,\tau')\delta_{D}(\kk-\kk'),
   \\
\langle\Phi_a(\kk,\tau)\Phi_b(\kk',\tau')\rangle 
      &= (2\pi)^3\delta_D(\kk+\kk')R_{ab}(|\kk|;\tau,\tau');\quad (\tau>\tau'). 
\end{align}
%
Then, the governing equations for power spectrum, equivalent to the 
renormalized expansions truncated at the one-loop level, become 
\cite{Taruya2008}
%
\begin{align}
\widehat{\Sigma}_{abcd}P_{cd}(k;\tau) &=
 \int^\tau_{\tau_0}d\tau''\,M_{as}(k;\tau,\tau'')R_{bs}(k;\tau,\tau'')
 \notag \\
 & +
 \int^\tau_{\tau_0}d\tau''\,N_{a\ell}(k;\tau,\tau'')G_{b\ell}(k|\tau,\tau'')
 \notag \\
&+ (a\leftrightarrow b),\label{eq:dP}\\
\widehat{\Lambda}_{ab}R_{bc}(k;\tau,\tau') &=
 \int^\tau_{\tau_0}d\tau''\,M_{as}(k;\tau,\tau'')R_{\overline{sc}}(k;\tau'',\tau') \notag \\
 & +  \int^{\tau'}_{\tau_0}d\tau''\,N_{a\ell}(k;\tau,\tau'')G_{c\ell}(k|\tau',\tau''),\label{eq:dR}\\
\widehat{\Lambda}_{ab}G_{bc}(k|\tau,\tau') &=
 \int^\tau_{\tau'}d\tau''\,M_{as}(k;\tau,\tau'')G_{sc}(k|\tau'',\tau').
\label{eq:dG}
\end{align}
%
Here, $R_{\overline{sc}}(k;\tau'',\tau')=R_{sc}(k;\tau'',\tau')$ for 
$\tau''>\tau'$, $R_{cs}(k;\tau',\tau'')$ for $\tau''<\tau'$. 
The operator $\widehat{\Sigma}_{abcd}$ is defined by
%
\begin{equation}
 \widehat{\Sigma}_{abcd}(\tau) = 
   \delta_{ac}\delta_{bd}\frac{\partial}{\partial\tau}
  +\delta_{ac}\Omega_{bd}(\tau) +\delta_{bd}\Omega_{ac}(\tau).
\end{equation}
%
The matrices $M_{ab}$ and $N_{ab}$ are 
%
\begin{align}
 M_{as}(k;\tau,\tau'') &= 4\int\frac{d^3\kk'}{(2\pi)^3}
   \gamma_{apq}(\kk-\kk',\kk')\gamma_{\ell rs}(\kk'-\kk,\kk) \notag \\
   &\times  G_{q\ell}(k'|\tau,\tau'')R_{pr}(|\kk-\kk'|;\tau,\tau''),
       \label{eq:def_M}\\
 N_{a\ell}(k;\tau,\tau'') &= 2\int\frac{d^3\kk'}{(2\pi)^3}
   \gamma_{apq}(\kk-\kk',\kk')\gamma_{\ell rs}(\kk-\kk',\kk') \notag \\
   &\times  R_{qs}(k';\tau,\tau'')R_{pr}(|\kk-\kk'|;\tau,\tau'').
       \label{eq:def_N}
\end{align}
%
Note that we have recast the original equations in Ref.~\cite{Taruya2008} 
in more symmetrical way by changing the integration variable 
[c.f. Eqs.~(49)--(53) of Ref.~\cite{Taruya2008}]. 
By definition, the non-linear propagator 
and the cross power spectra should satisfy the boundary condition:  
%
\begin{eqnarray}
 &&G_{ab}(k|\tau,\tau) = \delta_{ab},
 \label{eq:bc_G}
\\
 &&\lim_{\tau'\to\tau}R_{ab}(k;\tau,\tau') = P_{ab}(k;\tau).
 \label{eq:bc_R}
\end{eqnarray}
%

The closure equations (\ref{eq:dP})--(\ref{eq:dG}) are the integro-differential 
equations involving several non-linear terms, in which the   
information of the higher-order correction in SPT is encoded. Thus, 
replacing the statistical quantities $R_{ab}$ and $G_{ab}$ 
in non-linear terms with linear-order ones, the solutions of closure 
equations automatically reproduce the leading-order results of SPT, i.e., 
one-loop power spectra. Here, the linear-order quantities denoted by 
$R^{\rm L}_{ab}$ and $G^{\rm L}_{ab}$ satisfy 
%
\begin{align}
\widehat{\Lambda}_{ab}G^{\rm L}_{bc}(k|\tau,\tau') &= 0, \label{eq:dG3}\\
\widehat{\Lambda}_{ab}R^{\rm L}_{bc}(k;\tau,\tau') &= 0;\quad (\tau>\tau') .
  \label{eq:dR3}
\end{align}
%
For the rest of this paper, we focus on the numerical treatment of 
the closure equations and demonstrate the evolution 
of matter power spectrum in both non-linear and quasi-linear regimes by 
changing the treatment of non-linear terms.

\section{Numerical Method}
\label{sec:numerical}

The closure equations (\ref{eq:dP})--(\ref{eq:dG}) are the non-linear coupled 
equations involving the time-consuming integrals over space and time. 
In order to numerically treat these messy integrals while 
keeping computational cost, 
we implement the method used by Ref.~\cite{Valageas2007A}, in which the 
propagator and power spectra are expanded 
by a set of basis functions of $k$, and integrated
with respect to $k$ in advance of the time evolution. 
We adopt the trapezoidal rule for the integration with respect to $\tau$
and $k$, and the central difference formula for the time evolution. 
To be precise, we first prepare a discretised set of $k$ labeled as 
$k_m$ for $m=1,\cdots, N_k$, where we denote $k_1$ and $k_{N_k}$ by
$k_{\rm min}$ and $k_{\rm max}$, respectively.  
We define a set of triangular-shaped functions as the basis functions:
%
\begin{equation}
 \mathcal{T}_m(k) = \begin{cases}
\displaystyle
\frac{k-k_{m-1}}{k_m-k_{m-1}};&  k_{m-1} \leq k < k_m, \\
\displaystyle
\frac{k_{m+1}-k}{k_{m+1}-k_{m}};&  k_{m} \leq k < k_{m+1}, \\
\displaystyle
0;& {\rm otherwise}.
\end{cases} \label{eq:def_TM}
\end{equation}
Then we expand the non-linear propagator, the auto- and cross-power 
spectra as
%
\begin{align}
 P_{ab}(k';\tau) &= \sum_m\mathcal{P}_{ab,m}(\tau)\mathcal{T}_m(k'),
   \label{eq:expand_P} \\
 R_{ab}(k';\tau',\tau) &= \sum_m\mathcal{R}_{ab,m}(\tau',\tau)\mathcal{T}_m(k'),
   \label{eq:expand_R} \\
 G_{ab}(k'|\tau',\tau) &= \sum_m\mathcal{G}_{ab,m}(\tau',\tau)\mathcal{T}_m(k').
   \label{eq:expand_G} 
\end{align}
%
The above expressions together with basis function (\ref{eq:def_TM}) imply that 
the power spectra and the propagator between the discrete points are
evaluated by the linear interpolation according to the definition of the
basis functions. 
Note that these functions do not satisfy the orthogonality in the sense that  
the integration of the product $\mathcal{T}_m(k)\mathcal{T}_n(k)$ 
over the continuous space of $k$ does 
not vanish even if $m\neq n=m+1$. The set of $\mathcal{T}_m(k)$ 
has the orthogonality only on the discrete space because 
$\mathcal{T}_m(k_i)=\delta_{mi}$ is satisfied.

Substituting Eqs.~(\ref{eq:expand_P})--(\ref{eq:expand_G}) into 
Eqs.~(\ref{eq:def_M}) and (\ref{eq:def_N}), we obtain a separable form of the
matrices $M_{ab}$ and $N_{ab}$ :
%
\begin{align}
 M_{as}(k;\tau,\tau'') &= \sum_{\rm all\;indices}
   \mathcal{G}_{q\ell,m}(\tau,\tau'')
   \mathcal{R}_{pr,n}(\tau,\tau'')
   \mathcal{T}^{(M)}_{apq,\ell rs,m,n}(k), \label{eq:calcM} \\
 N_{a\ell}(k;\tau,\tau'') &= \sum_{\rm all\;indices} 
   \mathcal{R}_{qs,m}(\tau,\tau'')
   \mathcal{R}_{pr,n}(\tau,\tau'')
   \mathcal{T}^{(N)}_{apq,\ell rs,m,n}(k), \label{eq:calcN}
\end{align}
%
where $\mathcal{T}^{(M)}$ and $\mathcal{T}^{(N)}$ are given by
%
\begin{align}
\mathcal{T}^{(M)}_{apq,\ell rs,m,n}(k) &= 4\int\frac{d^3\kk'}{(2\pi)^3}
   \gamma_{apq}(\kk-\kk',\kk')\gamma_{\ell rs}(\kk'-\kk,\kk) \notag\\
    & \times \mathcal{T}_{m}(k')\mathcal{T}_{n}(k-k'),  \label{eq:TM}\\
\mathcal{T}^{(N)}_{apq,\ell rs,m,n}(k) &= 2\int\frac{d^3\kk'}{(2\pi)^3}
   \gamma_{apq}(\kk-\kk',\kk')\gamma_{\ell rs}(\kk-\kk',\kk') \notag\\
    &  \times \mathcal{T}_{m}(k')\mathcal{T}_{n}(k-k').  \label{eq:TN}
\end{align}
%
The details of the description on the integrations
(\ref{eq:TM}) and (\ref{eq:TN}) can be found in Appendix \ref{appendix:int}. 
Although Eqs.~(\ref{eq:TM}) and (\ref{eq:TN}) seem to have many 
components, most of them vanishes because the vertex function 
$\gamma_{abc}$ has
only three non-vanishing components given in Eq.~(\ref{eq:def_gamma}).
The relevant components in the summation are listed in Table~\ref{tab:aslist}. 
\begin{table}[!ht]
\centering
\begin{tabular}{c|l}\hline
$(a,s)$ & $(apq,\ell rs)$ for $M$ \\ \hline
(1,1) & (112,121), (121,121) \\
(1,2) & (112,112), (112,222), (121,112), (121,222) \\
(2,1) & (222,121) \\
(2,2) & (222,112), (222,222) \\ \hline\hline
 $(a,\ell)$ & $(apq,\ell rs)$ for $N$ \\ \hline
 (1,1) & (112,112), (112,121), (121,112), (121,121) \\
 (1,2) & (112,222), (121,222) \\
 (2,1) & (222,112), (222,121) \\
 (2,2) & (222,222) \\
\hline
\end{tabular}
\caption{List of non-vanishing components of the function
 $\mathcal{T}_{apq,\ell rs,m,n}^{(M,N)}(k)$ 
 in Eqs.~(\ref{eq:calcM}) and (\ref{eq:calcN}).} 
\label{tab:aslist}
\end{table}

\begin{figure}[t]
\centering{
  \includegraphics[width=12cm]{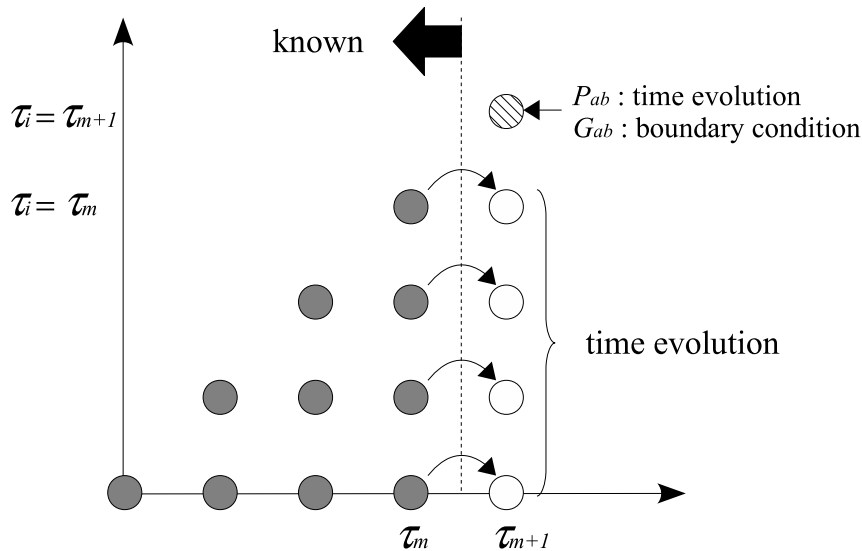} 
}
\caption{Schematic draw of the numerical procedure to solve 
closure equations. Each circle represents both $\mathcal{G}_{ab,p}$ and 
$\mathcal{R}_{ab,p}$ ($\mathcal{P}_{ab,p}$ for $\tau_i=\tau_m$) on the 
 discrete time grid, just omitting the wavenumber dependence. 
 The horizontal and vertical axes correspond to
 the second and third arguments of those quantities. 
 In the steps 3 and 4 mentioned in the main text, we compute 
 the matrices $M_{ab}(k;\tau_m,\tau_i)$ and $N_{ab}(k;\tau_m,\tau_i)$ 
 for $0\leq i\leq m$, and evaluate the non-linear terms, depicted 
 as filled circles at the right edge of the 'known' region. In next 
 step 5, we advance the time step and obtain 
 $\mathcal{G}_{ab,p}(\tau_{m+1},\tau_i)$ and 
 $\mathcal{R}_{ab,p}(\tau_{m+1},\tau_i)$ for 
 $0\leq i\leq m$ (blank circles), and $\mathcal{P}_{ab,p}$ and 
 $\mathcal{G}_{ab,p}$ for
 $(\tau_{m+1},\tau_b{m+1})$ (shaded circle).} 
 \label{fig:evolve}
\end{figure}

Based on the essential points of our numerical treatment described above, we 
now consider how to solve the closure equations in a cosmological setup. 
Basically, we perform the following steps (see also Fig.~\ref{fig:evolve}): 

\begin{enumerate}
\item Set $\tau=\tau_{\rm init}$ (or $z=z_{\rm init}$) for sufficiently
      small value of $\tau_{\rm init}<0$, where the universe is well-described 
      by the Einstein-de Sitter (EdS) model, and impose the initial
      conditions: 
%
\begin{equation}
 P_{ab}(k;\tau_{\rm init}) = R_{ab}(k;\tau_{\rm init},\tau_{\rm init}) 
    = \begin{pmatrix}1&1\\1&1\end{pmatrix}P_{\rm init}(k) ,
    \quad
 G_{ab}(k|\tau_{\rm init},\tau_{\rm init}) = \delta_{ab},
\label{eq:initial}
\end{equation}
%
      where $P_{\rm init}(k)$ is the linear power spectrum given at the initial
      time $\tau_{\rm init}$. Note that, in order to ensure the validity of 
      this prescription, the initial condition should be imposed early enough
      so that the influences of the non-linearity and the transient from 
      the initial condition can be neglected. Appropriate value of 
      the initial time has been chosen based on the convergence test 
      in Appendix \ref{appendix:convergence}. 
      
\item Perform the integration over $k$ for all components of 
  $\mathcal{T}^{(M)}_{apq,\ell rs,m,n}$ and 
  $\mathcal{T}^{(N)}_{apq,\ell rs,m,n}$ 
  using the trapezoidal rule [see Eqs.~(\ref{eq:TM}) and (\ref{eq:TN})].

\item Suppose that the coefficients $\mathcal{G}_{ab,p}(\tau_m,\tau_i)$  
    and $\mathcal{R}_{ab,p}(\tau_m,\tau_i)$ for $0\leq i\leq m$ have 
      been already obtained (filled circles in Fig.~\ref{fig:evolve}), 
      we perform the summation in Eqs.~(\ref{eq:calcM}) and (\ref{eq:calcN}), 
      and compute the kernels 
      $M_{ab}(k;\tau_m,\tau_i)$ and $N_{ab}(k;\tau_m,\tau_i)$. 

\item Calculate the non-linear terms in the right-hand side of 
      Eqs.~(\ref{eq:dP})--(\ref{eq:dG}). We use the trapezoidal rule for 
      the integration over time $\tau''$. 

\item Advance the time step from $\tau_m$ to $\tau_{m+1}$, 
    and obtain a new set of arrays 
      $G_{ab}(k|\tau_{m+1},\tau_i), R_{ab}(k;\tau_{m+1},\tau_i)$ 
      and $P_{ab}(k;\tau_{m+1})$ for $0 \leq i\leq m$, 
      applying the central difference formula to 
      Eqs.~(\ref{eq:dP})--(\ref{eq:dG}) 
      (open circles in Fig.~\ref{fig:evolve}). For the edge of 
      arrays (shaded circle in Fig.~\ref{fig:evolve}),  
      the boundary conditions 
      (\ref{eq:bc_G}) and (\ref{eq:bc_R}) are used to obtain 
      $G_{ab}(k|\tau_{m+1},\tau_{m+1})$ and 
      $R_{ab}(k;\tau_{m+1},\tau_{m+1})$.  

\item Repeat the steps 3--5 until the time $\tau_{n+1}$ reaches the 
    final time. 

\end{enumerate}

Note that the trapezoidal rule at step 4 and finite difference scheme 
at steps 5 are explicitly written as follows. 
For Eq.~(\ref{eq:dG}) with $(\tau,\tau')=(\tau_p,\tau_q)$, 
we have 
%
\begin{equation}
 \begin{aligned}
 &\frac{\mathcal{G}_{ab,r}(\tau_{p+1},\tau_q)- \mathcal{G}_{ab,r}(\tau_{p-1},\tau_q)}{2\Delta\tau}  
    + \Omega_{ab}(\tau_p)\mathcal{G}_{bc,r}(\tau_p,\tau_q)  \\
 &\quad\quad\quad\quad\quad\quad\quad\quad
 = \Delta\tau \sum_{m=q}^{p} c_m M_{as}(k_r;\tau_p,\tau_m) 
                                    \mathcal{G}_{sc,r}(\tau_m,\tau_q),
\end{aligned}\label{eq:diff_ex}
\end{equation}
%
for $p>q$. In cases with $p=q$, the differentiation in left-hand side of
the above equation is replaced with the first-order difference. 
Here, we set $c_m=1/2$ for $m=q,p$, otherwise $c_m=1$. 
Eqs.~(\ref{eq:dP}) and (\ref{eq:dR}) can be also written 
similarly as above.

The above procedure can be also used for
the calculation of the one-loop spectra in SPT. 
As we mentioned in Sec.~\ref{sec:closure}, 
the calculation in the SPT additionally needs the solutions 
$G^{\rm L}_{ab}$ and $R^{\rm L}_{ab}$ given in Eqs.~(\ref{eq:dG3}) and 
(\ref{eq:dR3}) in advance. They are 
used at the step 3 to compute the integration kernels,  
$M_{ab}^{\rm L}$ and $N_{ab}^{\rm L}$, which are evaluated from 
Eqs.~(\ref{eq:def_M}) and (\ref{eq:def_N}) 
just replacing the integrands with linear-order ones. 
In the same manner, at the step 4, we 
replace $G_{ab}$ and $R_{ab}$ in the non-linear interaction terms 
with linear-order quantities, $G_{ab}^{\rm L}$ and $R_{ab}^{\rm L}$, 
respectively.

\section{Demonstrations}
\label{sec:results}

In what follows, we present the results of numerical integration of 
closure equations. We first demonstrate the full non-linear 
calculation and present the results in Sec.~\ref{subsec:results_full}. 
Then, we move to discuss the perturbative treatment and examine the 
weakly non-linear evolution of the 
power spectrum in dark energy and modified 
gravity models in Sec.~\ref{subsec:results_perturbation}. 
The initial power spectrum $P_{\rm init}(k)$ is calculated from 
the linear transfer function in the flat $\Lambda$CDM model. We adopt 
the cosmological parameters determined from WMAP five-year results 
\cite{Komatsu2008}: 
$\Delta^2_{\mathcal{R}}(k_0=0.002\mbox{Mpc}^{-1}) = 2.457\times 10^{-9}$, 
$n_s=0.960$, $\Omega_{\rm m}=0.279$, $h=0.701$, for the amplitude of curvature
perturbation, scalar spectral index, density parameter of matter, and
Hubble parameter, respectively. Unless otherwise stated, we assume 
the dark energy with equation-of-state parameter $w_{\rm de}=-1$.

The parameters of our numerical calculations include  
the initial redshift $z_{\rm init}$, the cutoff wave number
$k_{\rm max}$, the number of time steps $N_\tau$, and the 
number of Fourier mesh $N_k$. We set $N_\tau=172$ and $N_k=200$ with 
constant interval in linear and logarithmic scales, respectively.  
For the initial redshift and cutoff wavenumber, 
based on the convergence test in Appendix~\ref{appendix:convergence}, 
we chose $z_{\rm init}=200$ and $k_{\rm max}=5\uk$. 
With this choice, the numerical errors in the SPT calculation are 
reduced to a sub-percent level.

\subsection{Full non-linear calculation}
\label{subsec:results_full}

\begin{figure}[t]
\centering
\includegraphics[width=12cm]{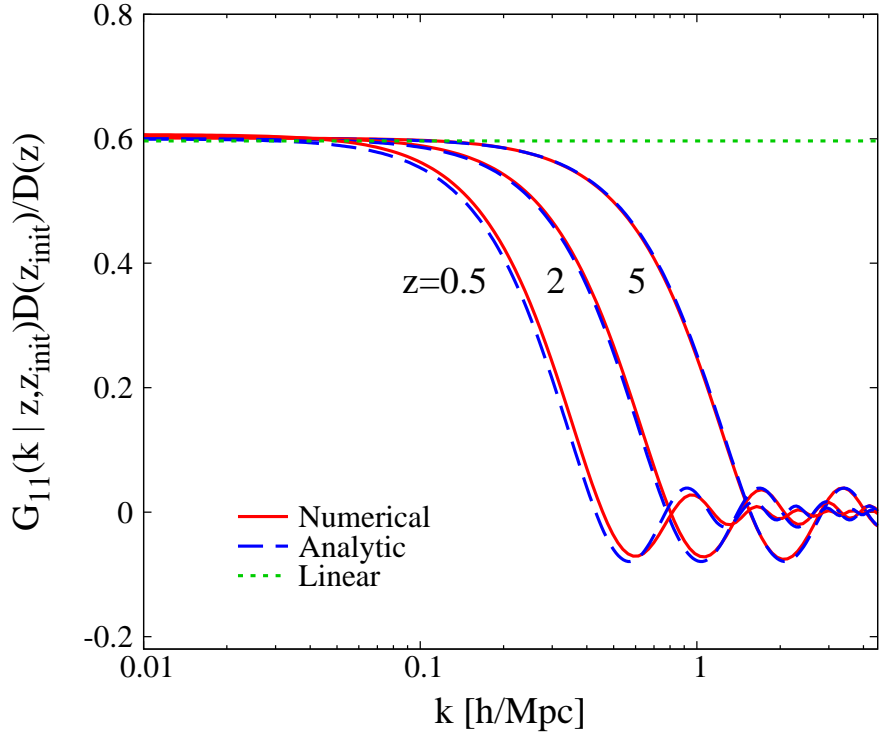} 
\caption{The non-linear propagators $G_{11}(k|z,z_{\rm init})$
 at $z=0.5, 2$ and $5$ from left to right. 
 The vertical axis is normalised by the linear growth rate, $D(z)$.
 The solid lines represent the numerical solution of the
 closure equations (\ref{eq:dP})--(\ref{eq:dG}).
 The dashed line are the approximate
 solution derived in Ref.~\cite{Taruya2008}, and the dotted line
 indicates the linear theory prediction.}
\label{fig:fullG}
\end{figure}
\begin{figure}[!ht]
\centering{
  \includegraphics[width=18cm]{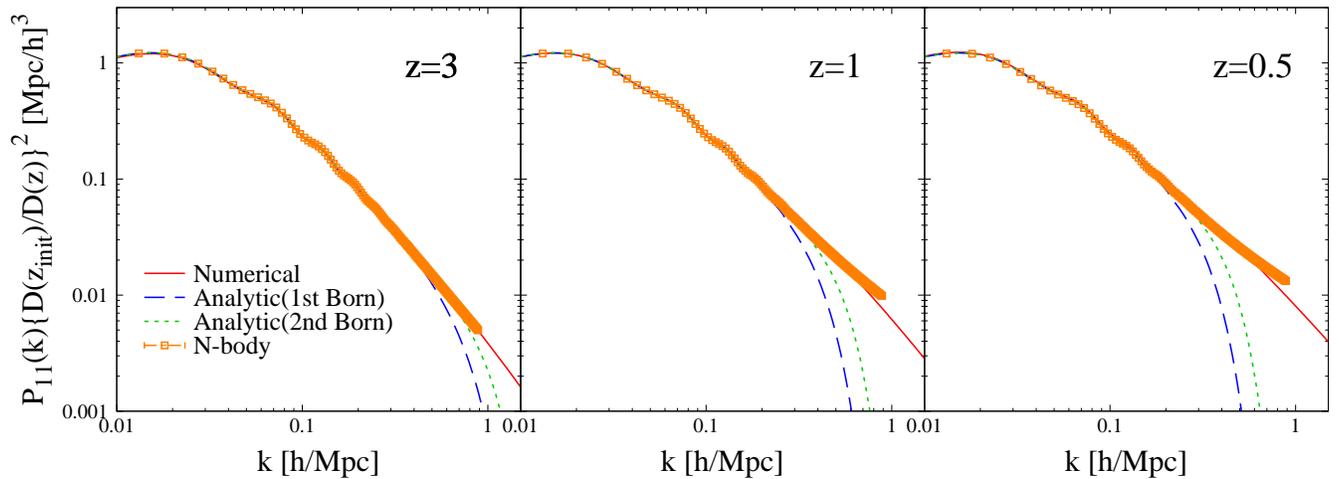} 
}
\caption{Time evolution of matter power spectra, $P_{11}(k)$, evaluated
 at $z=3,1$ and $0.5$ from left to right panels. 
 The vertical axis is normalised by the linear growth rate, $D(z)$.
 The solid lines represent
 the numerical solution of the 
 equations (\ref{eq:dP})--(\ref{eq:dG}). The dashed and
 dotted lines are the analytic results including the corrections up to
 the first- and second-order Born approximation, respectively \cite{Taruya2008}.
 The squares with errorbar indicate the N-body simulations
 taken from Ref.~\cite{Taruya2009}.}
\label{fig:full}
\end{figure}

In the full non-linear treatment, 
the solutions of 
auto- and cross-power spectra as well as the non-linear propagator are 
simultaneously obtained from the closure equations 
at each time step. Here, for illustrative purpose, 
we first show the non-linear propagators, which clearly manifest 
the non-perturbative property of non-linear clustering 
incorporated into our formalism.

Fig.~\ref{fig:fullG} plots the non-linear propagator 
$G_{11}(k|z,z_{\rm init})$ as function of wavenumber given at different 
redshifts, $z=0.5$, $2$ and $5$ (from left to right). 
Clearly, the numerical results depicted as solid lines
exhibit the damping oscillation, 
and asymptotically approach zero at $k\to\infty$. The characteristic scale of 
the damping is shifted to a lower $k$ for decreasing the redshift. 
These behaviors are marked contrast with the linear theory prediction 
depicted as dotted line. Note that the results including the 
leading-order correction (one-loop SPT) slightly 
improves the low-$k$ behavior, but  
they eventually become negative and diverge at $k\to \infty$. 
In this respect, the damping properties seen in the numerical results can be 
regarded as the non-perturbative effect, which results from 
the resummation of infinite series of higher-order corrections. 
Indeed, the damping behavior in the non-linear propagators 
has been already confirmed in the N-body simulations 
\cite{Crocce2005B,Bernardeau2008}, and is essential for 
the accurate prediction of power spectrum \cite{Crocce2007}.

In Fig.~\ref{fig:fullG}, the dashed lines indicate the analytic results 
obtained from Ref.~\cite{Taruya2008}. Basically, these are the approximate 
solutions of Eq.~(\ref{eq:dG}) constructed by matching the asymptotic 
solutions in the low-$k$ and high-$k$ limits. Although 
the analytic results at lower redshifts slightly deviate from the numerical 
solutions, the overall agreement between these two curves is remarkable. 
This may be an independent check for the stability of our numerical scheme, 
and 
the accuracy of our code seems comparable to or even better  than the
analytic calculations.

Now, in Fig.~\ref{fig:full}, we show the redshift evolution of the 
matter power spectrum, $P_{11}(k;z)$, obtained from the closure equations.  
For comparison, we also plot the N-body results taken from 
Ref.~\cite{Taruya2009}. 
Solid lines represents the numerical results of closure equations, 
dashed and dotted lines are the results of analytic 
calculations including up to the leading-order and next-to-leading order 
perturbative corrections, respectively. Here, 
the analytic results were obtained based on 
the integral solutions of the closure equations presented in 
Ref.~\cite{Taruya2008,Taruya2009}. We employ the Born 
approximation to evaluate the integral solutions perturbatively. 
Although the analytical treatment is found to accurately describe the 
non-linear evolution of baryon acoustic oscillations with 
a precision of sub-percent level \cite{Nishimichi2008,Taruya2009}, 
because of the perturbative treatment, 
applicable range of the analytic treatment is limited to 
a narrow range. As clearly shown in Fig.~\ref{fig:full}, 
the resultant power spectra rapidly fall off at some higher wavenumbers.
By contrast, the power spectra obtained from the numerical 
calculation first trace the analytical results on large scales, 
and they extend over small scales without a sharp drop of the amplitude. 
Remarkably, the numerical results quite resemble the N-body results 
at $k\lesssim1\,\,(0.6)h\,$Mpc$^{-1}$ for $z=3$ ($z=0.5$), and 
the agreement between these two results reaches the accuracy of 
$\sim4\% (8\%)$ level. 
This is a clear manifestation of the fact that
full non-linear treatment of the closure equations 
is indeed a non-perturbative way of calculating the power spectrum beyond 
the weakly non-linear regime. Hopefully, it would be 
a fast computational tool complementary to the N-body simulations. 
To clarify the usefulness of this approach, a more quantitative
comparison between N-body simulations and our numerical treatment is
needed. We will discuss this issue in a future work.

\subsection{Perturbative calculation} 
\label{subsec:results_perturbation}

In this subsection, we turn to focus on the perturbative treatment of 
the closure equations, by which all the quantities in non-linear terms 
are replaced with the linear-order ones. As we mentioned, 
this treatment automatically reproduces the one-loop results of SPT. 
Owing to the numerical treatment, we can address weakly non-linear 
evolution even when the analytical calculations are no longer possible. 
In Sec.~\ref{subsec:1loopPT_in_DE}, we discuss the one-loop power spectra 
in dark energy models, and address the validity of the analytical treatment 
based on the Einstein-de Sitter approximation. 
In Sec.~\ref{subsec:1loopPT_in_MG}, we examine 
a class of modified gravity models with linear Poisson equation, where 
the effective Newton constant manifestly depends on scale. We demonstrate 
how the modification of the gravitational-force law affects 
the power spectra in weakly non-linear regime.

%
\subsubsection{Dark energy models}
\label{subsec:1loopPT_in_DE}
%

\begin{figure}[t]
\centering{
  \includegraphics[width=8.5cm]{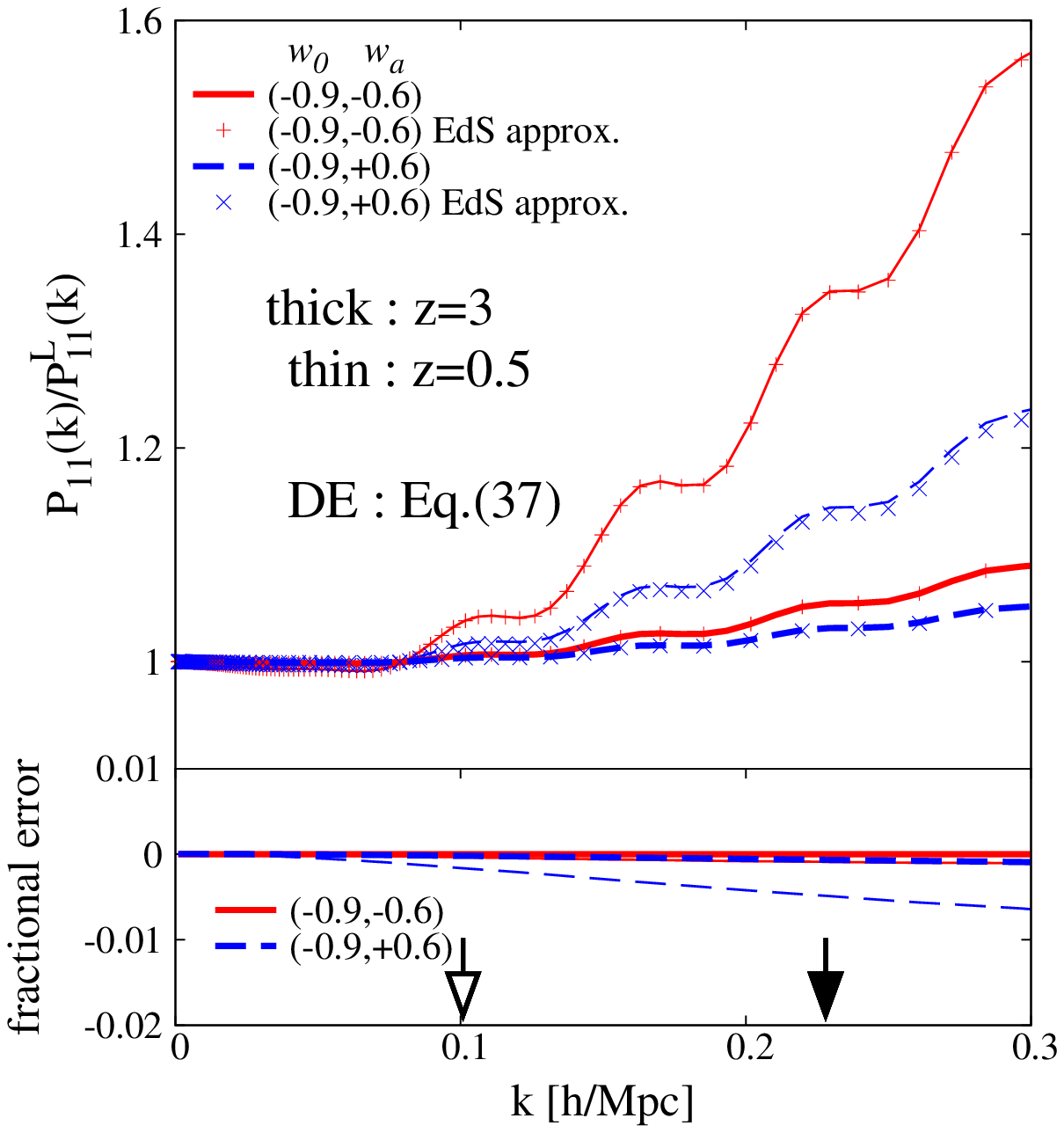}
  \includegraphics[width=8.5cm]{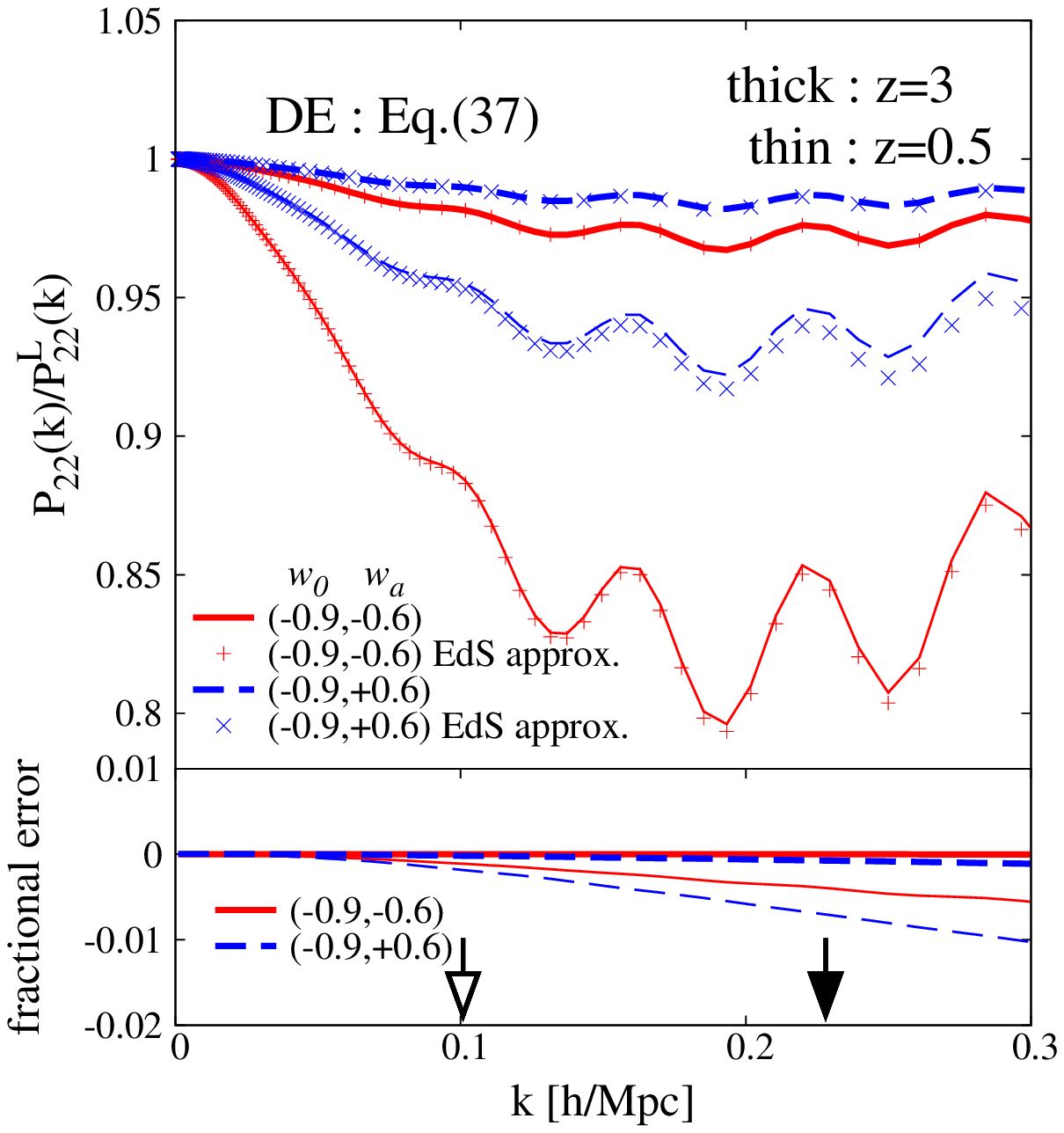}
}
\caption{The density ({\it left}) and velocity divergence power spectra
 ({\it right}) normalized by the linear theory predictions 
 in dark energy model with variable $w_{\rm de}$ [see 
 Eq.~(\ref{eq:var_w})]. The symbols and lines respectively represent the
 results with and without the EdS approximation:
 $(w_0,w_a)=(-0.9,-0.6)$ (solid lines and sumbol
 '$+$'); $(-0.9,+0.6)$ (dashed lines and symbol '$\times$'). In bottom panels,
 the differences between these results are plotted
 as the fractional error, $\{P^{\rm (EdS)}(k)-P^{\rm (num)}(k)\}/P^{\rm 
(num)}(k)$. 
 The thick (thin) lines are the results at $z=3 (0.5)$.
 The vertical arrows indicate the
 maximum wave number below which the prediction of SPT is expected to 
 agree well with the N-body simulations within $1\%$ (see
 Eq.~(\ref{eq:kc}), and Ref.~\cite{Nishimichi2008}).
 The filled and open arrows are the maximum wave numbers at $z=3$ and
 $z=0.5$, respectively.}
\label{fig:vw1}

\end{figure}
\begin{figure}[h]

\centering{
  \includegraphics[width=8.5cm]{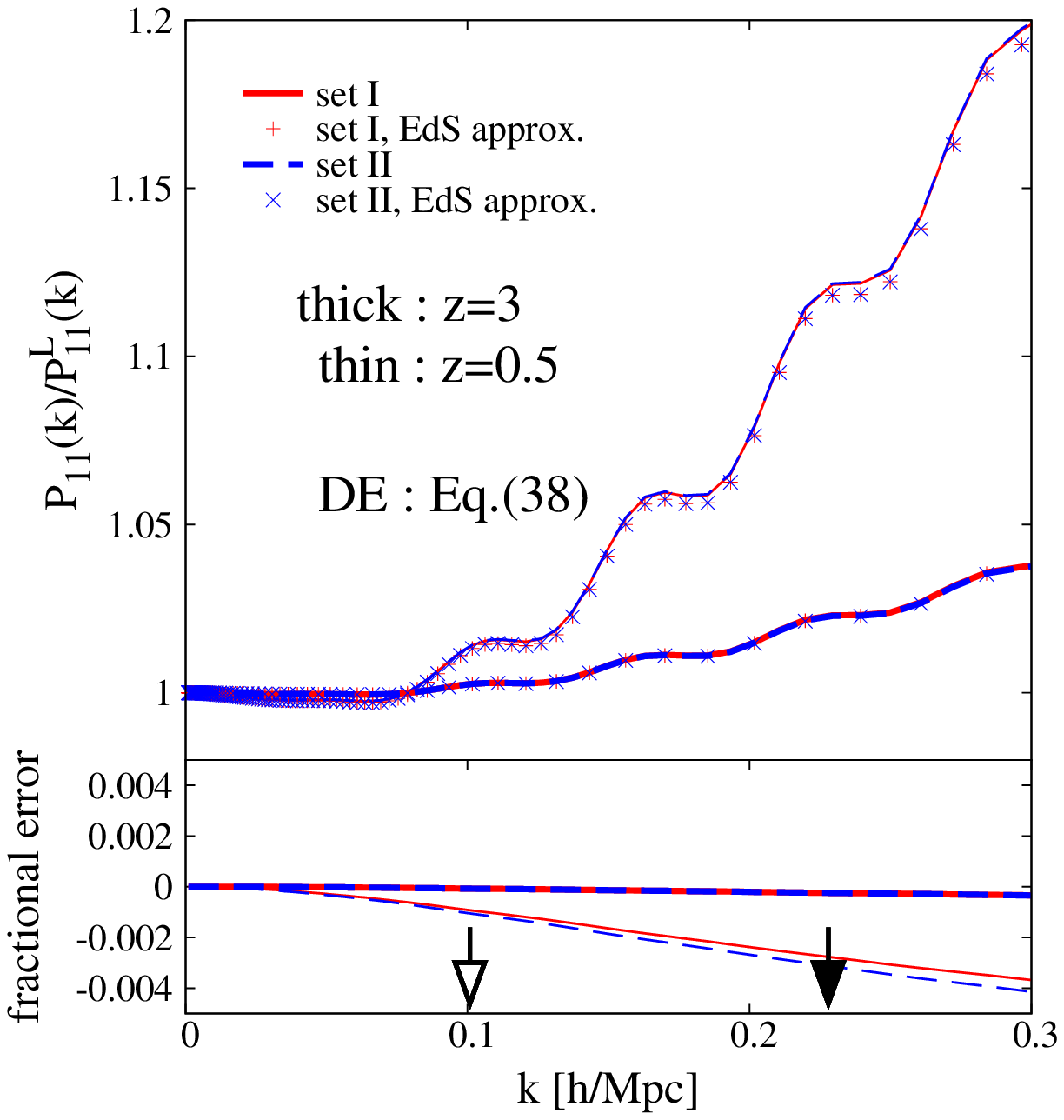}
  \includegraphics[width=8.5cm]{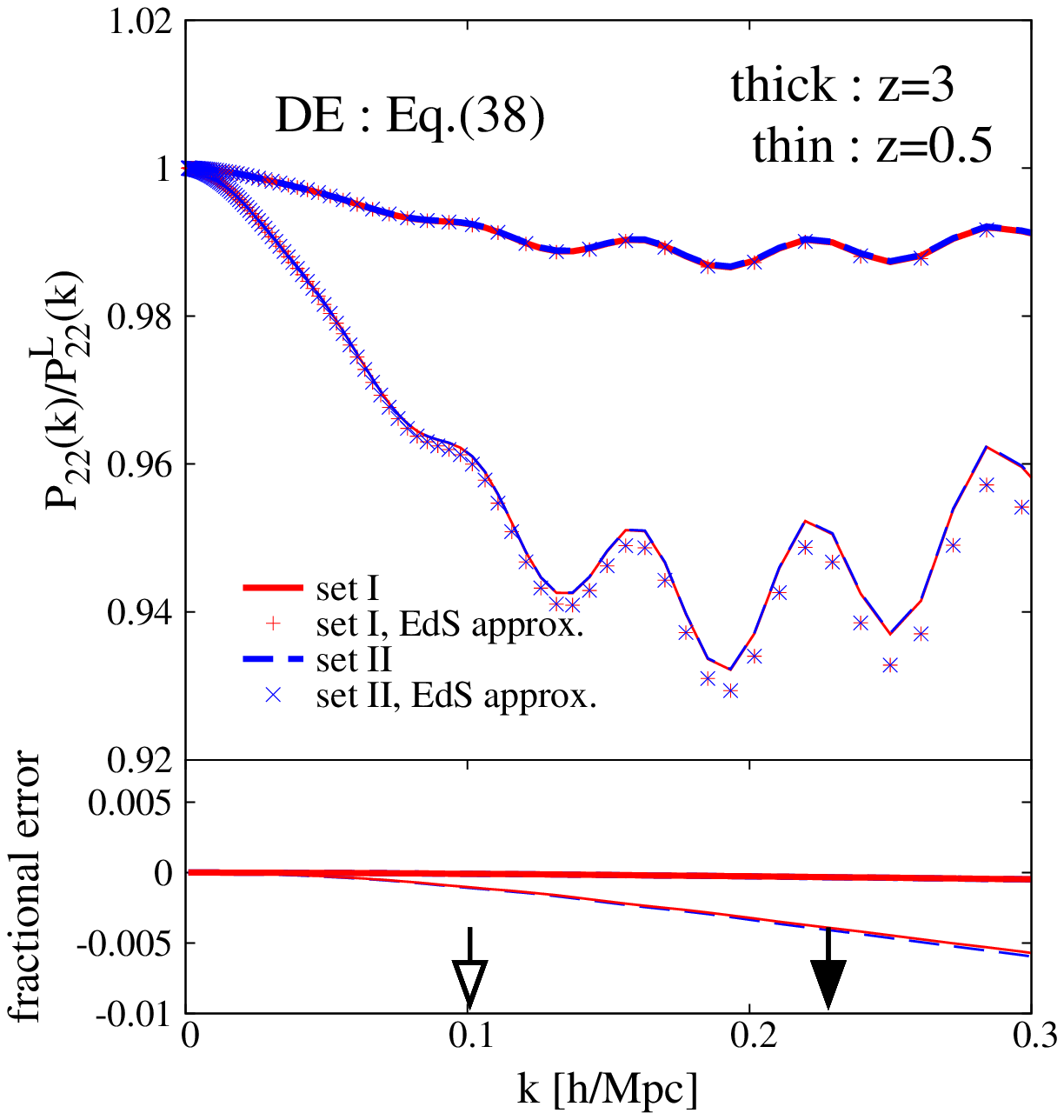}
}
\caption{Same as in Fig.~\ref{fig:vw1}, but in the dark energy model with 
  equation-of-state parameter (\ref{eq:w_HM}). 
  The model parameters are  set to 
 $(w_0,w_1,a_s,q)=(-1.8,-0.4,0.5,3.41)$ for {\it set I}, and 
 $(-1.8,-0.4,0.5,25.0)$ for {\it set II}. Note that the 
case with parameters of {\it set II} is regarded as extreme one, in which 
the effective equation-of-state parameter, 
$w_{\rm eff}\equiv-1-(2/3)d\ln H/d\tau$, sharply changes its sign at 
$a\sim0.5$. As for the vertical arrows, see the caption of Fig.~\ref{fig:vw1}.} 
\label{fig:hm}
\end{figure}

The one-loop SPT has recently attracted renewed interest for an 
accurate modeling of large-scale structure. In particular, 
a precise measurement of baryon acoustic oscillations made by 
ongoing and/or upcoming galaxy surveys to probe the nature of 
late-time cosmic acceleration provide a strong motivation to use 
the one-loop SPT for an accurate template of matter power spectrum 
(e.g., Refs.~\cite{Jeong2006,Nishimichi2007,Nishimichi2008,Jeong2009}). 
In these experiments, the required accuracy for theoretical template 
reaches at a percent level.

In the analytic treatment of one-loop power spectra, 
the Einstein-de Sitter (EdS) approximation has been frequently used in the 
literature (e.g., Ref.~\cite{Bernardeau2002} and references
therein). Under the approximation, the higher-order solutions of  
perturbation are approximately described by the linear growth 
factor $D(z)$, and the resultant power spectra are schematically 
expressed as  
%
\begin{equation}
 P_{ab}(k;z) = D^2(z)P_{ab}^{\rm L}(k) + D^4(z)P_{ab}^{\rm 1\mbox{-}loop}(k)+
\cdots. 
\label{eq:one-loop_Pk}
\end{equation}
%
Note that for dark energy models in general relativity, 
the EdS approximation is mathematically equivalent to solving 
the closure equations just replacing the matrix $\Omega_{ab}$ in the 
operators $\widehat{\Lambda}_{ab}$ and $\widehat{\Sigma}_{abcd}$ with
%
\begin{equation}
 \Omega_{ab}^{\rm EdS}(\tau) = 
\begin{pmatrix}
 0 && -1 \\
\\
{\displaystyle -\frac{3}{2}f^2} && 
{\displaystyle \frac{f}{2}-\frac{d\ln f}{d\tau} }
\end{pmatrix}, \label{eq:OmegaEDSA}
\end{equation}
%
with the function $f$ defined by $f\equiv d\ln D/d\tau$.

Here, we consider two specific examples of dark energy models 
characterized by the equation-of-state parameter $w_{\rm de}$ as 
\cite{Chevallier2000, Linder2002}
%
\begin{equation}
 w_{\rm de}(a) = w_0 + w_a(1-a), 
\label{eq:var_w}
\end{equation}
%
and \cite{Hannestad2004}
%
\begin{equation}
 w_{\rm de}(a) = w_0w_1\left(\frac{a^q + a_s^q}{w_1a^q + w_0a_s^q}\right).
 \label{eq:w_HM}
\end{equation}
%
Comparing the numerical results of closure equations with  
the analytical calculations, we discuss the validity of EdS approximation.

Fig.~\ref{fig:vw1} shows the one-loop spectra $P_{11}(k)$ 
({\it left}) and $P_{22}(k)$ ({\it right}) at $z=0.5$ and $3$, 
for dark energy model with slowly 
varying $w_{\rm de}$ [Eq.~(\ref{eq:var_w})]. The model parameters 
$w_0$ and $w_a$ were appropriately chosen within the currently 
constrained values of $|1+w_0|\lesssim0.1$ and $|w_a|\lesssim0.6$ (e.g.,
Ref.~\cite{Komatsu2008}). In upper panels, we 
plot the ratio of power spectra, $P_{ab}(k)/P_{ab}^{\rm L}(k)$, 
while in lower panels, we plot the fractional difference between the 
results with and without EdS approximation, i.e., 
$\{P^{\rm (EdS)}(k)-P^{\rm (num)}(k)\}/P^{\rm (num)}(k)$, where 
$P^{\rm (EdS)}$ and $P^{\rm (num)}$ are respectively obtained from 
the analytic and numerical calculations. Similarly, 
in Fig.~\ref{fig:hm}, we plot the results in the dark energy 
model (\ref{eq:w_HM}), 
in which the equation-of-state parameter has a sharp transition 
from $w_1$ to $w_0$ at the scale factor $a\sim a_s$ for a large $q$.

The resultant power spectra with EdS approximation 
underestimate the numerical results without EdS approximation 
in  both the density and velocity-divergence part of auto-power spectra. 
As decreasing the redshift, the deviation from numerical results 
becomes significant, but a level of discrepancy is not so large. 
These are consistent with the previous findings by 
Refs.~\cite{Takahashi2008,Pietroni2008}, from the analysis 
of matter power spectrum. In Figs.~\ref{fig:vw1} and \ref{fig:hm}, 
the vertical arrows indicate the maximum wave number below which the 
precision level of one-loop SPT is expected to be better than $1\%$. 
According to Ref.~\cite{Nishimichi2008}, this is 
determined by solving the following equation with respect to 
the wavenumber $k$: 
%
\begin{equation}
  \frac{k^2}{6\pi^2}\int_{0}^{k}P^{\rm L}_{11}(q;z)\,dq=0.18.
\label{eq:kc}
\end{equation}
%
Note that the maximum wavenumbers given above have been empirically derived 
by comparison between N-body simulations and theoretical predictions, and 
it seems rather conservative estimates compared to those previously proposed 
\cite{Jeong2006,Sefusatti2007,Matsubara2007}. 
Keeping the limitation of the one-loop SPT in mind, we confirm that the 
analytical treatment with EdS approximation is a quite good description of 
the one-loop power spectra and the accuracy of this treatment can reach 
a sub-percent level. This is even true for the model (\ref{eq:w_HM}) with 
the extreme parameter set, i.e., 
$(w_0,w_1,a_s,q)=(-1.8,\,\,-0.4,\,\,0.5,\,\,25.0)$, 
in which the effective equation-of-state parameter 
$w_{\rm eff}\equiv-1-(2/3)d\ln H/d\tau$, rather than $w_{\rm de}$, sharply 
changes its sign at $a\sim0.5$ and eventually approaches 
$w_{\rm eff}\sim 1$.

Therefore, as long as the dark energy models in general relatively 
are concerned, the analytical calculation with EdS approximation 
is very accurate treatment within the validity range of predictions, 
and it can give a fast computation of the weakly non-linear power spectrum.

%
\subsubsection{Modified gravity models}
\label{subsec:1loopPT_in_MG}
%

\begin{figure}[t]
\centering{
  \includegraphics[width=17cm]{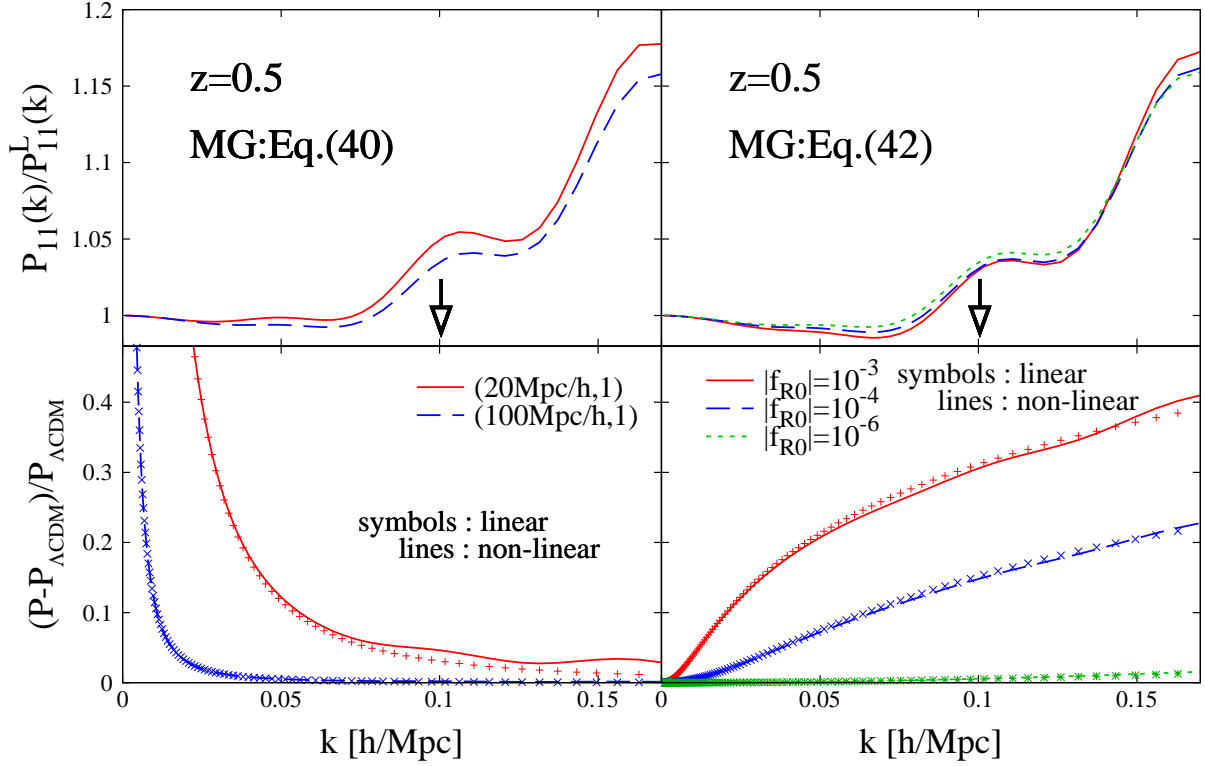} 
}
\caption{Weakly non-linear power spectrum at $z=0.5$ in a modified 
gravity model with Yukawa-type interaction ({\it left}) and $f(R)$ gravity 
model ({\it right}). The symbols, '$+$', '$\times$' and '$*$'
 respectively represent the resultant spectra
 in the linear theory, and lines the non-linear numerical solutions. The
 upper panels show the power spectra divided by the linear ones, and the
 lower ones indicate the deviations from the spectra calculated in the flat
 $\Lambda$CDM model. Note that the symbols in lower panels represent the
 linear theory predictions.
} 
 \label{fig:MG}
\end{figure}

Now let us consider the weakly non-linear evolution of the power spectrum 
in modified gravity models with linear Poisson equation. 
Unlike the dark energy models, the Newton 
constant is effectively modified, and even the linear growth rate generically 
depends on the scale and time. Thus, the EdS approximation cannot be 
applied in general and the analytical treatment is no longer 
possible\footnote{In the DGP model as one of the successful models that 
explains the late-time cosmic acceleration \cite{DGP}, 
the effective Newton constant $G_{\rm eff}$ depends only on time at 
the linear-order level, and the analytical calculation of one-loop 
spectrum is possible with a help of EdS approximation, 
even in the presence of non-linearity in Poisson 
equation \cite{Koyama2009}.}. Here, we demonstrate that 
with the use of the present formalism and numerical method,
the one-loop power spectrum can be accurately computed even in the 
analytically intractable cases.

We examine two representative modified gravity models 
whose effective Newton constant manifestly depends 
on the scale and time. One is a phenomenological model in which the 
Yukawa interaction is added by hand to the inverse-square law 
(e.g., Refs.~\cite{Sealfon2004, Shirata2005, Shirata2007}). The
effective Newton constant in this model is given by 
%
\begin{equation}
G_{\rm eff}(k,\tau)=G\,
\left\{1+\alpha\,\,\frac{1}{\lambda^{2}(k/a)^2+1}\right\}.
\label{eq:MNG}
\end{equation}
%
The parameter $\lambda$ is the characteristic (proper) length at which the 
Newton force is modified, and the amplitude $\alpha$ represents 
the strength of the deviation from the inverse-square law on large scales. 
Note that cosmological constraints on these parameters  
have been obtained recently from the galaxy power spectrum of 
the Sloan Digital Sky Survey \cite{Shirata2005, Shirata2007}. Based on
this, we compute the power spectra  for specific 
parameters  with $(\lambda,\,\alpha)=(20h^{-1}\,\mbox{Mpc},\,\,1)$ and 
$(100h^{-1}\,\mbox{Mpc},\,\,1)$.

As another example, we consider the $f(R)$ gravity model. This model 
has been recently attracted as one of the successful models that 
explains late-time cosmic acceleration
\cite{Hu2007A,Starobinsky2007,Appleby2007} (and see also
Ref.~\cite{Sotiriou2008} for a review). The  
$f(R)$ gravity model is given by the generalization of the
Einstein-Hilbert action to include arbitrary function of the scalar
curvature $R$:
%
\begin{equation}
 S = \int d^4x\sqrt{-g}\left[ \frac{R+f(R)}{16\pi G} + 
\mathcal{L}_{\rm m}\right],
\end{equation}
%
with $\mathcal{L}_{\rm m}$ being the Lagrangian of ordinary matter. Under the 
quasi-static treatment relevant for the scales well-inside the 
Hubble horizon, the effective Newton constant becomes (e.g., 
Ref.~\cite{Oyaizu2008B})
%
\begin{equation}
G_{\rm eff}(k,\tau) = G\, \left\{\frac{4}{3}
   - \frac{1}{3}\frac{\overline{\mu}^2}{(k/a)^2+\overline{\mu}^2}\right\}, 
  \label{eq:Poisson_fr}
\end{equation}
%
where the quantity $\overline{\mu}^2$ is the effective mass of the new scalar 
degree of freedom, $f_R$, and is defined by 
$\overline{\mu}^2 = \{(1+f_R)/(\partial f_R/\partial R)-R\}/3$. 
Note that the barred quantity $\overline{\mu}^2$ implies the one evaluated in 
terms of the background quantities. In general, 
the corrections coming from non-linear interaction terms appear in 
$G_{\rm eff}$, but we do not consider here. In the present paper, 
we specifically consider the function $f(R)$ of the form, 
$f(R)\propto R/(A\,R+1)$ \cite{Hu2007A,Oyaizu2008A,Oyaizu2008B,Schmidt2008}. 
In the cosmologically 
interesting setup with $R\to0$ and $f(R)\to0$, this can be expanded as
%
\begin{equation}
f(R)\simeq-16\pi\,G\,\rho_{\Lambda} -f_{R0}\,
\left(\frac{\overline{R}_0}{R}\right).
\label{eq:fR_expand}
\end{equation}
%
The energy density $\rho_{\Lambda}$ is related with the constant $A$, and 
$\overline{R}_0$ and $f_{R0}$ are the background curvature and the field 
value given by $f_R(R_0)$ at present time. Here, we consider the 
cases with $|f_{R0}|\ll1$,  in which the last term at the right-hand side of 
Eq.~(\ref{eq:fR_expand}) is safely negligible and the background expansion just 
follows the same expansion history as in the $\Lambda$CDM model.

Fig.~\ref{fig:MG} shows the numerical results of one-loop power spectra 
given at $z=0.5$. Left and right panels plot the results for the models with 
effective Newton constant (\ref{eq:MNG}) and (\ref{eq:Poisson_fr}), 
respectively. The upper panels show the ratio of matter power spectrum, 
$P_{11}(k)/P_{11}^{\rm L}(k)$, while in lower panels, the fractional 
enhancement relative to the $\Lambda$CDM model, 
i.e., $\{P(k)-P_{\Lambda{\rm CDM}}(k)\}/P_{\Lambda{\rm CDM}}(k)$, 
is plotted. In model with Eq.~(\ref{eq:MNG}),  the modification of 
the gravitational-force law appears on large scales, and the effective Newton 
constant becomes $G_{\rm eff}\to (1+\alpha)\,G$.  On the other hand, the 
Newton constant on small scales becomes $4/3$ times greater than that on
large scales in the model with Eq.~(\ref{eq:Poisson_fr}). 
This scale-dependent nature qualitatively explains the results seen 
in the lower panels, and because of this, the resultant shape of the one-loop 
spectra is significantly altered. Nevertheless, when normalized by the 
linear power spectra, which intrinsically possesses the scale-dependent 
nature of $G_{\rm eff}$ through the linear growth rate, the differences 
in the mode transfer efficiency between two models turn out to be small 
(see upper panel). This indicates 
that the modification of gravitational-force law imprinted in the 
linear power spectrum can be preserved in the weakly non-linear regime, 
and the linear growth rate becomes an important clue to distinguish 
between various modified gravity models. This would be even true for a large 
class of the modified gravity models with non-linear Poisson equation.

Finally, it is interesting to note that in the 
model with Eq.~(\ref{eq:Poisson_fr}), there appears the crossing point 
at which the dependence of the ratio $P(k)/P^{\rm L}(k)$ on $|f_{R0}|$ is 
changed. As shown in the upper-right panel, 
the ratio decreases with $|f_{R0}|$ on large scales, while 
it eventually increases on small scales. This behavior basically 
reflects the fact that the strong gravity on small scales 
efficiently promotes the mode transfer from the low-$k$ to high-$k$ modes.

\section{Discussion and Conclusion}
\label{sec:conclusion}

In this paper, on the basis of the non-perturbative framework of 
the cosmological perturbation theory developed by Ref.~\cite{Taruya2008},  
we have presented a numerical scheme to treat non-linear 
evolution of matter power spectrum. The governing equations for matter 
power spectra are a closed set of evolution equations coupled with 
non-linear propagator, which has been previously derived by truncating 
the infinite chain of moment equations, with a help of perturbative 
calculation called closure approximation. The present formulation is 
equivalent to the one-loop level of renormalized perturbation theory, and 
the non-perturbative effects of gravitational clustering are effectively 
incorporated into the solution of closure equations. Note that 
the closure equations consistently reproduces the so-called 
one-loop results of standard perturbation theory if we replace the 
quantities in the non-linear terms with linear-order ones. 
The numerical scheme presented here can be used for 
the predictions of matter power spectra in both quasi non-linear and 
non-linear regimes, and is applicable to the analytically intractable 
cases. The modification to the gravity sector is straightforward.

We have demonstrated that the full non-linear treatment of the closure 
equations has a ability to treat non-linear evolution of power 
spectrum beyond the validity regime of previous analytical calculations. 
The resultant shape of the non-linear spectrum resembles the 
N-body result, and the agreement between these two results reaches 
the accuracy of $\sim4\% (8\%)$ level at $z=3(z=0.5)$.
We then focused on 
the perturbative treatment of closure equations, and presented the 
numerical results of one-loop SPT in various situations. 
We discussed the validity of the analytical treatment based on the 
Einstein-de Sitter approximation which has been frequently used in the 
literature. In the dark energy models with two representative 
equation-of-state parameters (\ref{eq:var_w}) and (\ref{eq:w_HM}),  
we found that the analytical calculation with Einstein-de Sitter 
approximation provides an excellent description for the density 
and velocity-divergence components of the one-loop power spectrum. 
Within the validity range of one-loop spectra, the accuracy of this 
treatment reaches at a sub-percent level. Also, we have studied 
the one-loop power spectra in a class of modified gravity models, 
in which the effective Newton constant manifestly depends on the 
scale and time, and the analytical calculation is no longer possible. 
We demonstrated that the scale-dependent modification of the 
gravitational-force law alters the power spectrum significantly, but the 
efficiency of the mode transfer arising from the non-linear mode coupling 
changes 
only moderately. In this respect, the modification of the gravity 
imprinted in the linear power spectrum would be 
preserved in the weakly non-linear regime, and the (scale-dependent) 
linear growth rate may be an important clue to distinguish 
between various modified gravity models.

The numerical scheme presented here is a first step toward 
precisely modeling the non-linear evolution of matter power 
spectrum in various situations.  
Recently, the non-linear spectrum including the massive neutrinos 
has been investigated by Ref.~\cite{Lesgourgues2009} 
based on the approach similar to our formalism
\cite{Pietroni2008}. Incorporating 
the effect of massive neutrinos into the present formalism is 
rather straightforward, and the closure equations may be used 
for a non-perturbative calculation of matter power spectrum 
beyond the free-streaming scales. As another direction, 
one may consider the extension of the present formulation to 
deal with a wide class of modified gravity models with non-linear 
Poisson equation. Ref.~\cite{Koyama2009} presented a general 
formalism to treat such models and explicitly calculated the one-loop 
power spectrum in DGP and $f(R)$ gravity models from the 
closure equations. The results for full non-linear treatment 
are left for future work, and will be reported elsewhere.

\begin{acknowledgments}
We would like to thank Takahiro Nishimichi for providing us the
numerical data of his N-body simulations. AT is supported
by a Grant-in-Aid for Scientific Research from the Japan Society for 
the Promotion of Science (JSPS) (No.~21740168). This work was 
supported in part by 
Grant-in-Aid for Scientific Research on Priority Areas No.~467
``Probing the Dark Energy through an Extremely Wide and Deep Survey with
Subaru Telescope'', and JSPS Core-to-Core Program ``International
Research Network for Dark Energy''.
\end{acknowledgments}

\appendix

\section{Details of numerical integrations in Eqs.(\ref{eq:TM}) and (\ref{eq:TN})}
\label{appendix:int}

In this appendix, we discuss the technical details on the numerical 
integrations of the non-linear terms in closure equations. 
In the numerical algorithm presented in Sec.~\ref{sec:numerical}, 
we must evaluate Eqs.~(\ref{eq:TM}) and (\ref{eq:TN}) in
advance to the time evolution. To compute these integrals, 
the expressions are first rewritten with the form of 
the two-dimensional integral with a help of the symmetry in the  
integrands. Then, we introduce the elliptic coordinate used in
Ref.~\cite{Valageas2007A}, and perform the integration 
by the trapezoidal rule taking carefully account of the domain of integration. 

\begin{figure}[!ht]
 \begin{minipage}{5cm}
   \includegraphics[width=5cm]{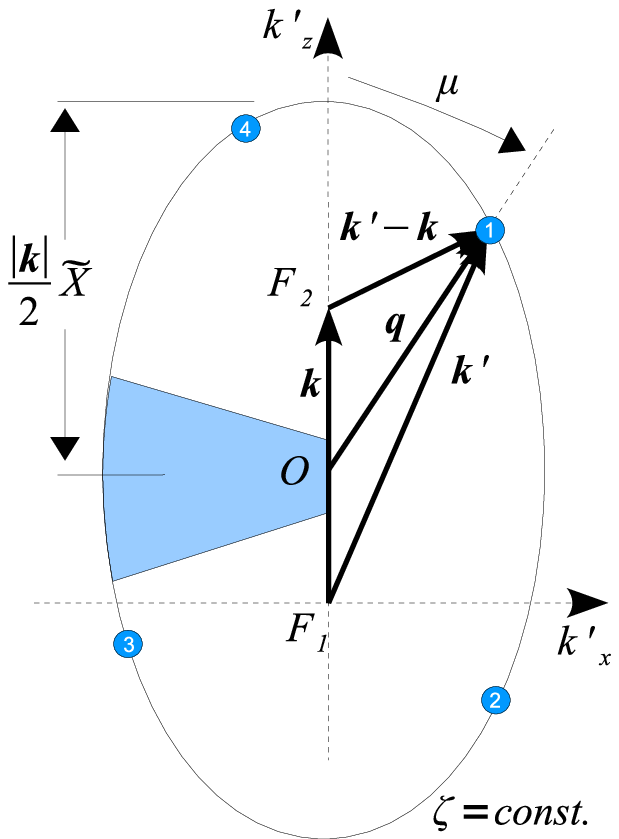} 
 \end{minipage}
   \hspace{1cm}
 \begin{minipage}{7cm}
   \includegraphics[width=7cm]{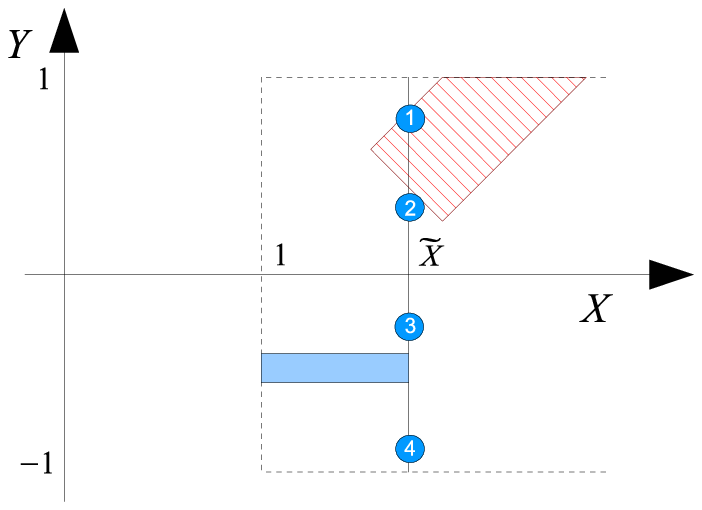} 
 \end{minipage}
 \caption{The geometrical relation between $\kk'$ (or $\qq$) space and
 $(X,Y)$ coordinates. The ellipic contour representing a surface with
 $\zeta=$ const. in  
 left plot is mapped to $X=\widetilde{X}$ in right plot. Circled
 numbers in both plots corresponds to each
 other, and filled areas, too. Note that the origin of $\kk'$-space is
 $F_1$, while that of the elliptic coordinate is $O$. The striped
 deficient rectangular is an actual domain defined by
 Eqs.~(\ref{eq:Y1})--(\ref{eq:X0}).}
 \label{fig:q-coord}
\end{figure}

The elliptic coordinate is defined as
%
\begin{equation}
\kk' = \frac{\kk}{2} + \mathbf{q},\qquad \mathbf{q}=\frac{k}{2}
 \begin{pmatrix}
   \sinh\zeta \sin\mu \cos\phi \\
   \sinh\zeta \sin\mu \sin\phi \\
   \cosh\zeta \cos\mu
 \end{pmatrix},
\end{equation}
%
where the vector $\kk$ is set to be aligned to the third axis. We introduce
%
\begin{equation}
 X=\cosh\zeta,\quad Y=\cos\mu, \label{eq:def_XY}
\end{equation}
%
where $X \geq 1$ and $-1 \leq Y \leq 1$. 
In Fig.~\ref{fig:q-coord}, we schematically plot the elliptic coordinate.
In left plot, the origin of the elliptic coordinate is $O$, and 
the elliptic contour represents a surface of $\zeta=$const, which is
mapped to $X=\widetilde{X}$ shown in right plot. 
The vector $\kk$ points from $F_1$ to $F_2$, which correspond to 
the two foci of the ellipse. Thus an arbitrary vector $\kk'$ and
$\kk-\kk'$ can be represented by the vector $\qq$.

In the elliptic coordinate, 
the vertex functions defined in Eq.~(\ref{eq:def_gamma}) are recasted as
%
\begin{align}
 \gamma_{112}(\kk-\kk',\kk') &= \frac{1+XY}{(X+Y)^2},
 &\gamma_{112}(\kk'-\kk,\kk)&=\frac{1+XY}{4},\\
 \gamma_{121}(\kk-\kk',\kk') &= \frac{1-XY}{(X-Y)^2},
 &\gamma_{121}(\kk'-\kk,\kk)&=\frac{X^2+Y^2-2}{2(X-Y)^2},\\
 \gamma_{222}(\kk-\kk',\kk') &= \frac{2(2-X^2-Y^2)}{(X^2-Y^2)^2},
 &\gamma_{222}(\kk'-\kk,\kk)&=\frac{XY-1}{4}\left(\frac{X+Y}{X-Y}\right)^2.
\end{align}
%
From now on, we take $(X,Y,\phi)$ as the integration variables instead
of $(k_x',k_y',k_z')$. Hence the volume element $d^3\kk'$ and the
integration domain are changed as
%
\begin{equation}
 \int_{|\kk'|=k_{\rm min}}^{|\kk'|=k_{\rm max}}\frac{d^3\kk'}{(2\pi)^3}(\cdots)
   = \frac{k^3}{32\pi^2}\int^{X_1}_{X_0}dX\int^{Y_1(X)}_{Y_0(X)}dY (\cdots),
 \label{eq:reduced}
\end{equation}
%
Note that, since the integrands of Eqs.~(\ref{eq:TM}) and (\ref{eq:TN}) are
axially symmetric, we can integrate over the azimuthal angle $\phi$, yielding
a factor $2\pi$. The lower and upper limits of the integration are
determined from the domains of definition of Eqs.~(\ref{eq:def_XY})
and (\ref{eq:def_TM}), which gives
%
\begin{align}
 Y_1(X) &= \min\left\{ -X+\frac{2k_{m+1}}{k}, X-\frac{2k_{n-1}}{k}, 1
 \right\}, \label{eq:Y1} \\
 Y_0(X) &= \max\left\{ -X+\frac{2k_{m-1}}{k}, X-\frac{2k_{n+1}}{k}, -1
 \right\}, \\
 X_1 &= \max\left\{ 1, \frac{k_{m+1}+k_{n+1}}{k}\right\}, \\
 X_0 &= \max\left\{ 1, \frac{k_{m-1}+k_{n-1}}{k}\right\}, \label{eq:X0}
\end{align}
%
where $k_{-1}$ and $k_{N+1}$ are assigned to $k_0=k_{\rm min}$ and
$k_N=k_{\rm max}$, respectively. With the above preparation, the
three-dimensional integration (\ref{eq:TM}) and (\ref{eq:TN}) are reduced to
two-dimensional integrations over the domain
(\ref{eq:Y1})--(\ref{eq:X0}). An example of integration domain is
depicted as a shaded deficient rectangle in right plot of
Fig.~\ref{fig:q-coord}.
As mentioned in Sec.~\ref{sec:numerical}, we implement the trapezoidal
rule to integrate Eqs.~(\ref{eq:TM}) and (\ref{eq:TN}) in this domain.

\section{Convergence test}
\label{appendix:convergence}

In this appendix, we check the convergence of numerical results
obtained with the numerical scheme mentioned in
Sec.~\ref{sec:numerical}. The test calculations have been done in the
$\Lambda$CDM model by varying some numerical parameters. Particularly we
focus on the initial time of the time evolution, $z_{\rm init}$, and the cutoff
wave-number on small scales, $k_{\rm max}$, introduced in
(\ref{eq:def_TM}), which are most sensitive parameters to the final
results. 

\begin{figure}[!ht]
\centering
\includegraphics[width=10cm]{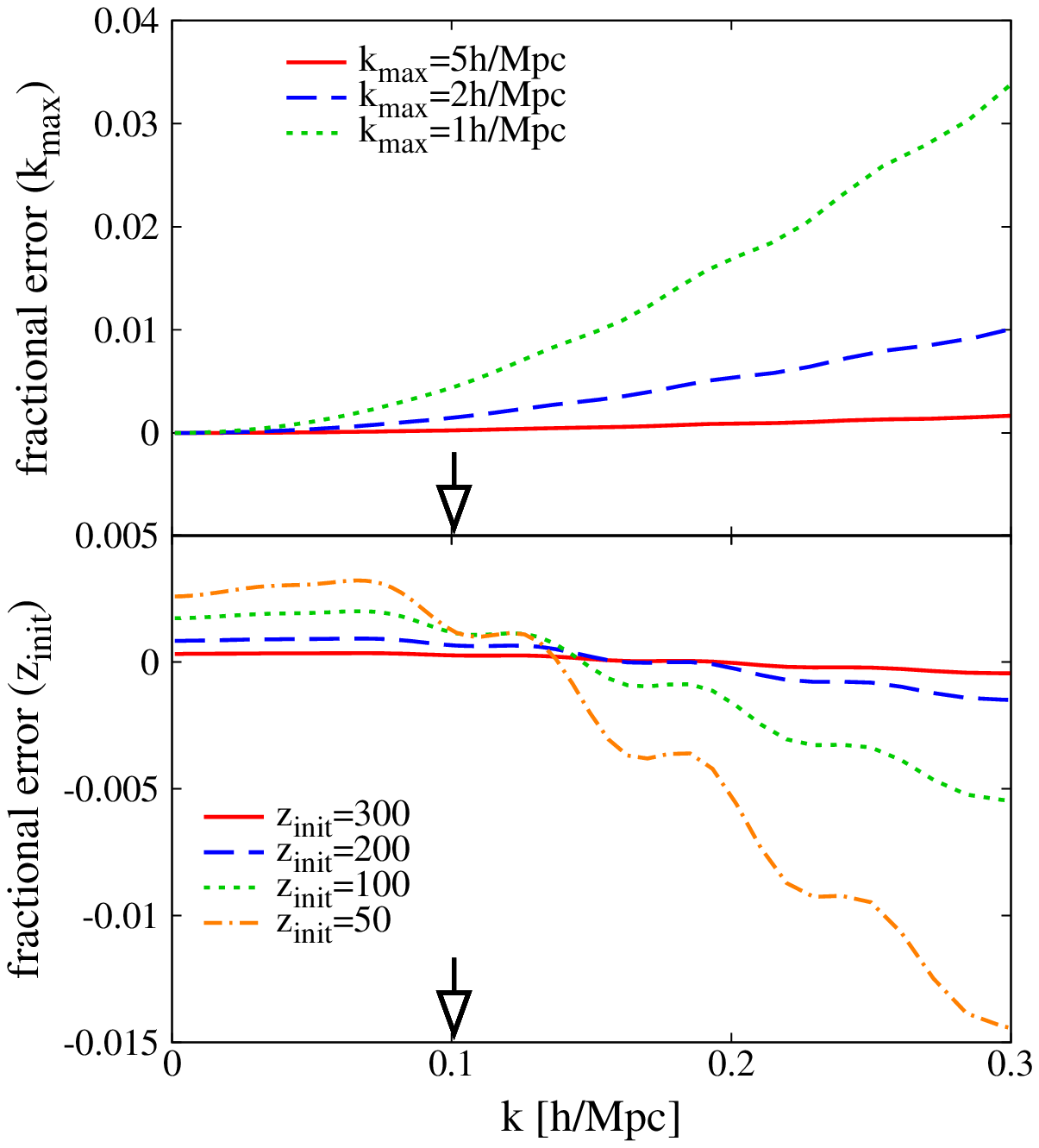} 
\caption{The dependence of the cutoff wave number, $k_{\rm max}$, and
 the initial time, $z_{\rm init}$, on the matter power spectrum. {\it
 Upper panel} : The fractional errors
 of the power spectrum for $k_{\rm max}=5\uk$ (solid), 
 $k_{\rm max}=2\uk$ (dashed) and $k_{\rm max}=1\uk$ (dotted) by
 reference to $k_{\rm max}=10\uk$. {\it Lower panel} : 
 The fractional errors for $z_{\rm init}=300$ (solid), 
 $z_{\rm init}=200$ (dashed), $z_{\rm init}=100$ (dotted)
 and $z_{\rm init}=50$ (dot-dashed) by reference to 
 $z_{\rm init}=400$. As for the vertical arrows, see the caption of
 Fig.~\ref{fig:vw1}.}  
\label{fig:numeric}
\end{figure}

The upper panel in Fig.~\ref{fig:numeric} shows the fractional
errors between the matter power spectrum $P_{11}(k)$ for
$k_{\rm max}=1,2,5\uk$ and that for $k_{\rm max}=10\uk$ denoted by
$P_{11}^{\rm ref}$, that is,
$(P_{11}-P^{\rm ref}_{11})/P^{\rm ref}_{11}$.
In these calculations, the logarithmic interval $\Delta(\log k)$ is fixed.
In the lower panel, we show the fractional errors for the calculations
with $z_{\rm init}=50-300$ from the one with $z_{\rm init}=400$. Also in these
calculations, we fixed the time interval, $\Delta\tau$. The arrows on
the horizontal axis are given by Eq.~(\ref{eq:kc}). 

Both plots indicate the good convergence of the numerical results in the
sense that the fractional errors from each reference become smaller as
$k_{\rm max}$ and $z_{\rm init}$ increase.
We found that we can keep the fractional error sufficiently
smaller than $1\%$ as long as $k_{\rm max}$ is larger than 5$\uk$, and
$z_{\rm init}$ is larger than 200. Particularly, as for $z_{\rm init}$, if we take later
time, the resultant power spectrum at low-redshifts
is harmed by the fact that we neglect the decaying modes in the
initial conditions [see Eq.~(\ref{eq:initial})].

Considering the above results, we fixed $k_{\rm max}=5\uk$ and
$z_{\rm init}=200$ for all numerical calculations presented in this
paper. Additionally the number of time steps, $N_\tau$, and the wave
number bins, $N_k$, are chosen as $N_\tau=172$ and $N_k=200$,
respectively, so that the fractional errors are suppressed to a
sub-percent level. Moreover, for the integration (\ref{eq:reduced}), we
use a $200\times 200$ discrete grid on the integration domain defined by
Eqs.~(\ref{eq:Y1})--(\ref{eq:X0}).

%
%

\bibliographystyle{apsrev}

\end{document}